\documentclass[conference]{IEEEtran}
\usepackage{amsmath,amssymb,dsfont,stfloats,color,url,mathtools}
\usepackage[pdftex]{graphicx}
\usepackage{wasysym,esint}
\usepackage{cite}
\usepackage{caption}
\usepackage{subcaption}
\usepackage{filecontents}

\usepackage{gensymb} 
\usepackage[boxruled,linesnumbered]{algorithm2e}
\usepackage{algorithm2e}
\usepackage{units}
\usepackage{lipsum}

\DeclareGraphicsExtensions{.eps,.pdf,.png,.jpg,.gif,.jpeg,.pstex}

\newfont{\bbb}{msbm10 scaled 700}

\newfont{\bb}{msbm10 scaled 1100}
\newcommand{\CC}{\mbox{\bb C}}

\newcommand{\av}{{\bf a}}
\newcommand{\bv}{{\bf b}}

\newcommand{\Sm}{{\bf S}}
\newcommand{\Tm}{{\bf T}}

\usepackage{stmaryrd} 

\usepackage{hyperref}
\hypersetup{
    bookmarks=true,         
    unicode=false,          
    pdftoolbar=true,        
    pdfmenubar=true,        
    pdffitwindow=false,     
    pdfstartview={FitH},    
    pdfnewwindow=true,      
    colorlinks=true,        
    linkcolor=red,          
    citecolor=green,        
    filecolor=blue,         
    urlcolor=blue           
}

\title{Array-Fed RIS: Validation of Friis-Based Modeling Using Full-Wave Simulations}

\author{\IEEEauthorblockN{Krishan K. Tiwari$^1$, Thomas Flisgen$^{2,3}$, Wolfgang Heinrich$^{1,2}$, Giuseppe Caire$^{1}$}
	\IEEEauthorblockA{$^1$Technische Universit\"at Berlin, 10587 Berlin, Germany.~~~$^2$Ferdinand-Braun-Institut (FBH), 12489 Berlin, Germany.\\~~~$^3$Brandenburgische Technische Universität Cottbus - Senftenberg, 03046 Cottbus, Germany.\\
		Email addresses: {lastname}@tu-berlin.de, {firstname.lastname}@b-tu.de, {firstname.lastname}@FBH-Berlin.de
  }}
\begin{document}

\bstctlcite{BSTcontrol} 

\maketitle

\begin{abstract}

Space-fed large antenna arrays offer superior efficiency, simplicity, and reductions in size, weight, power, and cost (SWaP-C) compared to constrained-feed systems. Historically, horn antennas have been used for space feeding, but they suffer from limitations such as bulky designs, low aperture efficiency ($\approx50\%$), and restricted degrees of freedom at the continuous aperture. In contrast, planar patch arrays achieve significantly higher aperture efficiency ($>90\%$) due to their more uniform aperture distribution, reduced weight, and increased degrees of freedom from the discretized aperture. Building on these advantages, we proposed an array-fed Reflective Intelligent Surface (RIS) system, where an active multi-antenna feeder (AMAF) optimizes power transfer by aligning with the principal eigenmode of the AMAF-RIS propagation matrix $\Tm$. While our previous studies relied on the Friis transmission formula for system modeling, we now validate this approach through full-wave simulations in CST Microwave Studio®. By comparing the Friis-based matrix, $\Tm_{\rm Friis}$, with the full-wave solution, $\Tm_{\rm full.wave}$, we validate the relevance of the Friis-based modeling for top-level system design. Our findings confirm the feasibility of the proposed AMAF-RIS architecture for next generation communication systems.


\end{abstract} 

\begin{IEEEkeywords}
Array-fed reflectarrays, reflective intelligent surface (RIS), full-wave simulations, Friis-based modeling validation, D-band communications.
\end{IEEEkeywords}

\section{Introduction}  
\label{sec:intro}

Large antenna arrays and higher carrier frequencies are likely to play pivotal roles in future communication systems. In fact, large antenna arrays have already been central to satellite communications and even radio sensing for several decades now. It is well known that space feeding large apertures enables higher hardware efficiency, simplicity, reliability, and reduced size, weight, power, and cost (SWaP-C) compared to guided-wave feeding \cite{2018MRA}. For legacy reasons (starting from (before) Berry's 1963 reflectarray \cite{1963} when microstrip patch antenna hardware was not realized), horn has been used as the primary antenna for space feeding. For an optimal horn, the aperture efficiency is about 50$\%$ only. Recall that the aperture efficiency of a planar patch array is much higher ($> 90\%$) due to a more uniform aperture distribution \cite[Fig. 6]{Munson74} than that of ``legacy'' feed horns. This is helpful to alleviate the aperture blockage problem, more so with easier mounting mechanism for a light-weight, low-profile patch array than the bulky, cumbersome horn mounting mechanism.

With the above motivation, we had proposed passive reflectarrays (a.k.a. reflective intelligent surface (RIS)) fed by an active multi-antenna feeder (AMAF), where the AMAF is configured along the principal eigenmode (PEM) of the AMAF-RIS propagation matrix $\Tm$ for maximum power transfer \cite{VTC2022, ICC2023, VTC2024, NewOldIdea, SPAWC}. There we used the Friis transmission formula and the distance dependent phase term based AMAF-RIS propagation matrix $\Tm_{\rm Friis} \in \CC^{N_p \times N_a}$ given by
\begin{equation} 
T_{k,\ell, \rm Friis}  = \frac{\sqrt{E_A(\varphi_{k,\ell},\vartheta_{k,\ell})E_R(\varphi_{k,\ell},\vartheta_{k,\ell}})}{2\pi r_{k,\ell}}~ e^{ - j\pi r_{k,\ell}},  \label{Telement}
\end{equation}
where 
$r_{k,\ell}$ is the distance (normalized by half wavelength) between 
the $k$-th RIS element and the $\ell$-th AMAF element, $(\varphi_{k,\ell},\vartheta_{k,\ell})$ denote the azimuth and elevation angles at which they see each other with respect to their own normal (boresight) direction. The patch element antenna (both at the RIS and the AMAF) is modeled 
by the classical axisymmetric power radiation pattern, \cite[Eq. (2-31)]{Balanis_antenna_theo}, \cite[Eq. (17)]{Pozar_1997}, 
\begin{equation} 
G_{\rm patch}(\varphi,\vartheta)=5.8\,\cos^2(\varphi)\,\cos^2(\vartheta),\label{Telement}
\end{equation}
the half-power beamwidth (HPBW) is 32.8$^\circ$, and the element gain is $\unit[7.64]{dBi}$. In other words, $E_A(\varphi,\vartheta) = E_R(\varphi,\vartheta) = G_{\rm patch}(\varphi,\vartheta)$. In \cite{NewOldIdea} we had observed that the principal singular value $\sigma_1>1$, however the AMAF-RIS propagation is a passive mechanism which cannot generate power. Furthermore, for the half-wavelength element spacing, we had been assuming negligible mutual coupling in our earlier works.

In order to assess and quantify the impact of the above ``idealized'', mutual-coupling agnostic, Friis-based modeling vis-a-vis a more realistic hardware conditions, we performed full-wave simulation based investigations in CST Microwave Studio\textregistered. At every element of the AMAF as well the RIS a port was placed and S-parameters\footnote{Recall that scattering parameters (S-parameters) \cite{S‐Parameters} define the input-output relationships between two ports under impedance matched condition at both the ports, and at high frequencies, power waves are used for analysis.}\cite{S‐Parameters} from the full-wave simulations were used to obtain the transmission coefficient (the communication theory canonical complex baseband channel coefficient) between every AMAF and every RIS element. 
In fact, the AMAF-RIS transmission coefficient matrix $\Tm$ can be considered as a block submatrix of the commonly known (in electromagnetics and RF network theory) scattering matrix \cite{S‐Parameters}:
\begin{equation}
\underbrace{\begin{pmatrix}
\ldots & \Tm\\
\ldots & \ldots
\end{pmatrix}}_{\Sm}
\underbrace{\begin{pmatrix}
\av_{\mathrm{RIS}}\\
\av_{\mathrm{AMAF}}
\end{pmatrix}}_{\av}=
\underbrace{\begin{pmatrix}
\bv_{\mathrm{RIS}}\\
\bv_{\mathrm{AMAF}}
\end{pmatrix}}_\bv\textnormal{,}\label{eq_smatrix}
\end{equation}
whereas
$\Sm\in\CC^{(N_p+N_a) \times (N_p+N_a)}$ is the scattering matrix of the structure. The vectors $\av_{\mathrm{RIS}}\in\CC^{N_p}$ and $\av_{\mathrm{AMAF}}\in\CC^{N_a}$
represent the incident wave amplitudes at the RIS and AMAF ports, respectively, while $\bv_{\mathrm{RIS}}\in \CC^{N_p}$ and $\bv_{\mathrm{AMAF}}\in \CC^{N_a}$ denote the corresponding outgoing wave amplitudes.
It follows from (\ref{eq_smatrix}) that the propagation matrix $\Tm \in \CC^{N_p \times N_a}$ characterizes the relationship between waves incident on the AMAF ports and those scattered at the RIS ports.

Thus, we obtained the full-wave solution based AMAF-RIS matrix $\Tm_{\rm full.wave}$. We study and compare the properties of $\Tm_{\rm full.wave}$ against $\Tm_{\rm Friis}$. We then use $\Tm_{\rm full.wave}$ in our MATLAB\textregistered\ communication system model with the 3D geometry \cite{SPAWC} to compare cumulative rate distribution (CDF) functions for the information theoretic communication rate using $\Tm_{\rm full.wave}$ and $\Tm_{\rm Friis}$. We chose this hybrid approach because it was shown in \cite{2013array} that an array (the RIS in our case) radiation performance characterization using array theory closely matches the results from aperture theory (unless very fine accuracies in far-off sidelobes and cross-polarization values are required, which is not the focus of this paper), where full-wave simulations had shown that both methods can be used equivalently for a reliable calculation of the general pattern shape, main beam direction, beamwidth, and sidelobe levels. Therefore, we choose the array theory based RIS radiation pattern calculation, which is faster and requires much less computation. 

The rest of the paper is organized as follows: Section \ref{sec:element} presents the patch element design used for both the AMAF and the RIS. Section \ref{sec:full_designs} presents the three dimensional AMAF-RIS model. In Section \ref{sec:CA} we analyze and compare $\Tm_{\rm full.wave}$ and $\Tm_{\rm Friis}$, evaluating their impact on communication system performance. Finally, Section \ref{sec:CONC} concludes the paper by summarizing the key highlights and outlines future research directions. 

\section{Patch Element Design} 
\label{sec:element}

\begin{figure}[h] 
\centerline{\includegraphics[width=3.5cm]{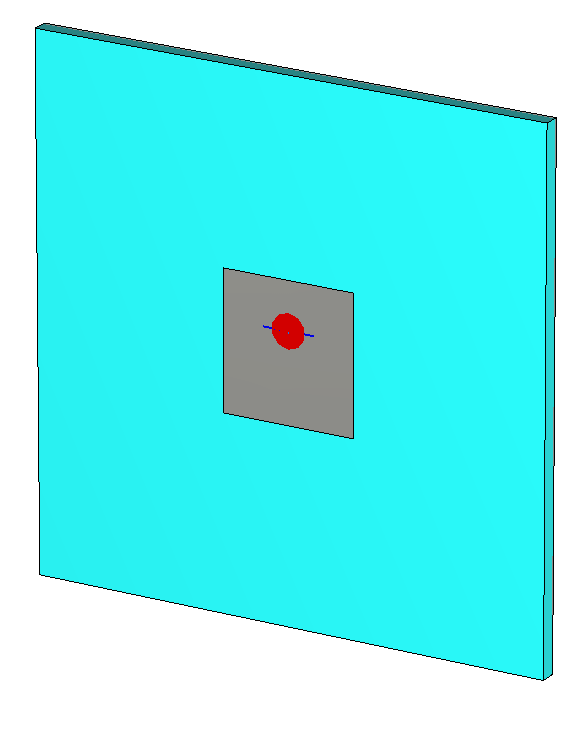}}
\caption{Patch Element connected to a discrete port highlighted in red. The position of the discrete port is shifted $\unit[80]{\mu m}$ from the patch center to ensure a base impedance close to $\unit[50]{\Omega}$.}
\label{fig:element}
\end{figure}

\begin{figure}[h] 
\centerline{\includegraphics[width = 0.95\linewidth]{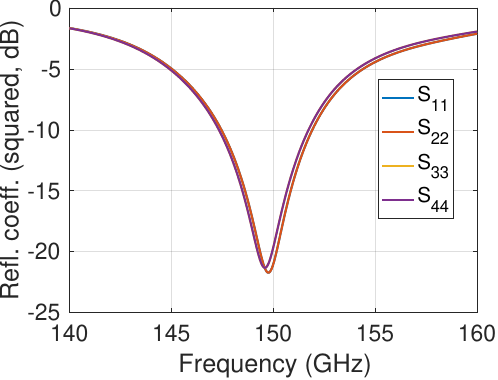}}
\caption{Reflection at individual patch elements (4 AMAF elements).}
\label{fig:reflection}
\end{figure}

Based on patch antenna element design principles \cite{PozarPatchBook}, a D-band microstrip patch element was designed at the center frequency of $\unit[150]{GHz}$. The thickness of the substrate is chosen to be $\unit[64]{\mu m}$ and its relative dielectric constant to be $3$. The width and the height of the individual patch are $\unit[514]{\mu m}$ and $\unit[527]{\mu m}$, respectively. Discrete ports are connected to each patch with an offset of $\unit[80]{\mu m}$ from its center, to obtain a base impedance whose real part is closed to $\unit[50]{\Omega}$. Fig.~\ref{fig:element} depicts one patch element with its discrete port. The gain of the patch element is $\unit[7.6]{dBi}$ and the squared absolute value of the reflection coefficient is shown in Fig.~\ref{fig:reflection}. 

\section{AMAF-RIS Design} 
\label{sec:full_designs}

\begin{figure}[h] 
\centerline{\includegraphics[width=4cm]{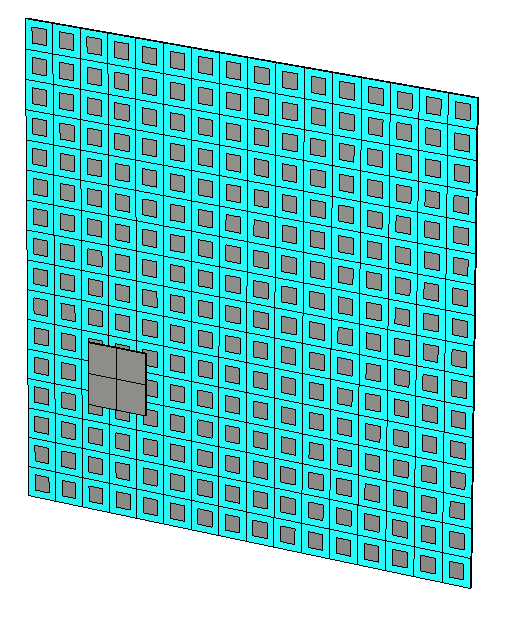}}
\caption{The 3D AMAF-RIS model in CST Microwave Studio\textregistered. The AMAF with $N_a=4$ patches is visible from the back and the RIS with $N_p = 256$ patches is visible from the front.}
\label{fig:3D_AMAF_RIS_CST}
\end{figure}

Based on the results from \cite{SPAWC}, using CST Microwave Studio\textregistered, we modeled a $16 \times 16$ standard rectangular array (inter-element spacing of half-wavelength) of the patch element in Section \ref{sec:element} as the RIS and a $2\times 2$ standard rectangular array as the AMAF. The AMAF-RIS focal distance was chosen to be $8$ half-wavelengths for an optimum $F/D=0.5$ \cite{ICC2023, NewOldIdea}. Discrete ports were placed at all $260$ patch elements, additional space was considered around the geometry, and open boundaries were specified so that (almost) no reflections from the boundary of the computational domain arise. In this way, the RIS and its AMAF depicted in Fig.~\ref{fig:3D_AMAF_RIS_CST} are modeled to exist in free space. Additionally, a frequency-domain field monitor at $\unit[150]{GHz}$ was defined to record the absolute value of the electric field strength on the surface of the RIS substrate.

The resulting problem is electrically large and $138{,}170{,}580$ hexahedral meshcells were used to discretize the depicted geometry. The time domain solver of CST Microwave Studio\textregistered~was used to one after the other excite the four patches of the AMAF with a Gaussian pulse carrying root mean square energy from $\unit[140]{GHz}$ to $\unit[160]{GHz}$. The stopping criterion for the time-domain iteration was set to $\unit[-50]{dB}$. Based on the time-domain excitation and the corresponding response, reflection and transmission factors in frequency domain are obtained by means of a Fast Fourier Transformation. In total, the computation of the response of the four excitations required more than $\unit[18]{h}$ on an Intel(R) Xeon(R) CPU E5-2620 v4 @ $\unit[2.10]{GHz}$ equipped with $\unit[128]{GB}$ RAM. Note that this approach does not provide the complete scattering matrix \eqref{eq_smatrix} at all. However, in this study, we are only interested in the transmission matrix $\Tm$. The fact that all relevant transmission coefficients are available from the successive excitation of the four AMAF patches lowers the computational burden.

\begin{figure}[h] 
\centerline{\includegraphics[width = 0.95\linewidth]{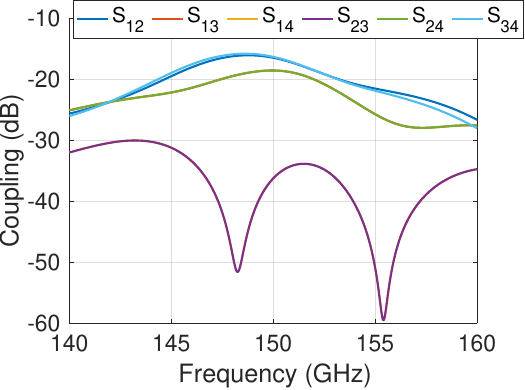}}
\caption{Mutual coupling between AMAF elements.}
\label{fig:coupling}
\end{figure}

The mutual coupling among AMAF elements is plotted in Fig. \ref{fig:coupling}, where we see that the inter-element mutual coupling is non-negligible. While the coupling between diagonally opposite elements is less than -30 dB (which is expected because of a larger diagonal distance), the coupling between adjacent elements is as high as -16 to -18 dB. While there are techniques to reduce inter-element mutual couplings, including choosing a high $\varepsilon_\mathrm{r}$ substrate and also increasing the inter-element spacing, we consider the current design with the half-wavelength inter-element spacing for maximum space utilization and also to investigate the robustness of Friis-based modeling against practical RF conditions with non-negligible mutual coupling.

\section{Comparative Analysis} 
\label{sec:CA}

The principal singular values $\sigma_{1\rm Friis}$ and $\sigma_{1\rm full.wave}$ for the $\Tm_{\rm Friis}$ and $\Tm_{\rm full.wave}$ are $1.6$ and $0.9$, respectively. Clearly, the $\sigma_{1\rm Friis}>1$ is the limitation of the ``simplistic'' Friis-based modeling.
In such scenarios, we recommend ceiling $\sigma_1$ to
unity and proceeding with the analysis for a high-level system
design because the performance impact of this approximation
is not dramatic as we will see later in Figs. \ref{fig:Gemic_pointing_error} and \ref{fig:Gemic_digital_PS}.


\begin{figure}[h] 
\centerline{\includegraphics[width=7.25cm]{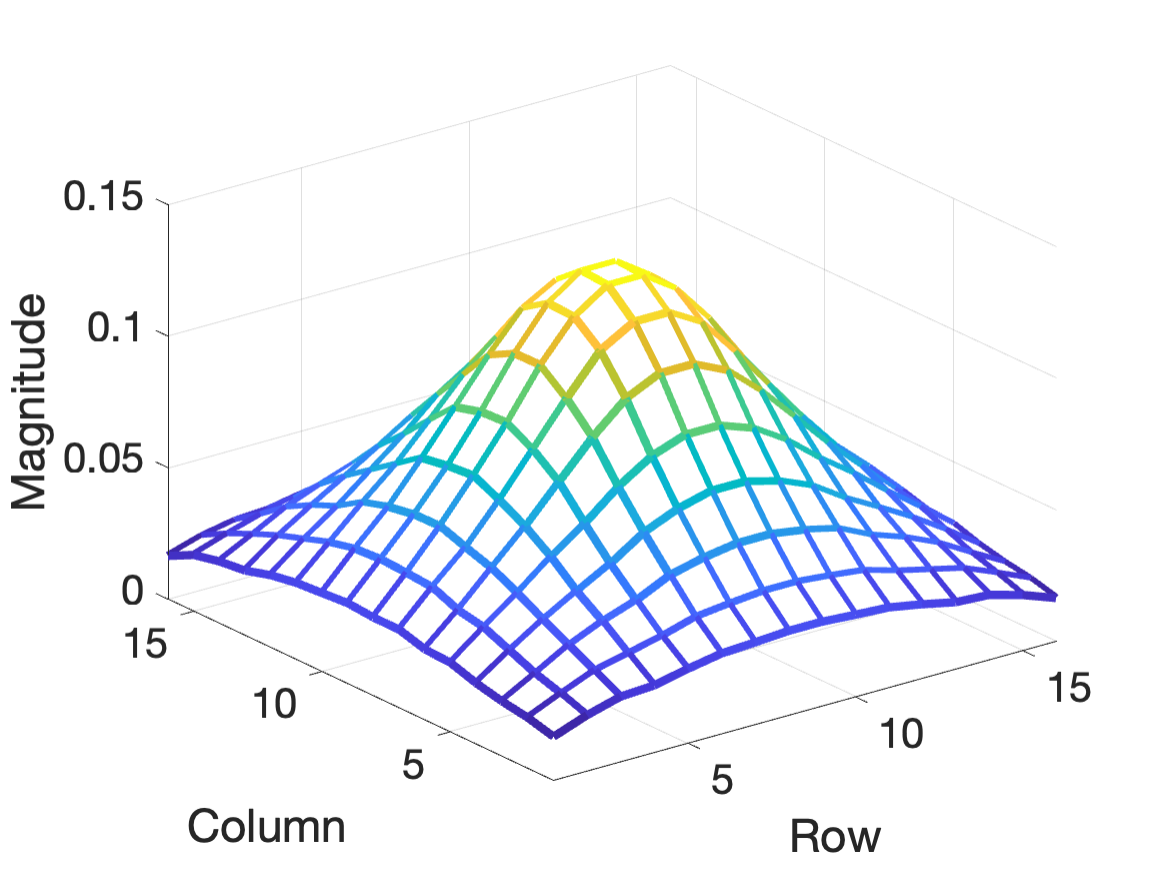}}
\caption{RIS PEM magnitude taper (normalized) as obtained from full-wave simulations.}
\label{fig:RIS_FW}
\end{figure}

\begin{figure}[h] 
\centerline{\includegraphics[width=7.25cm]{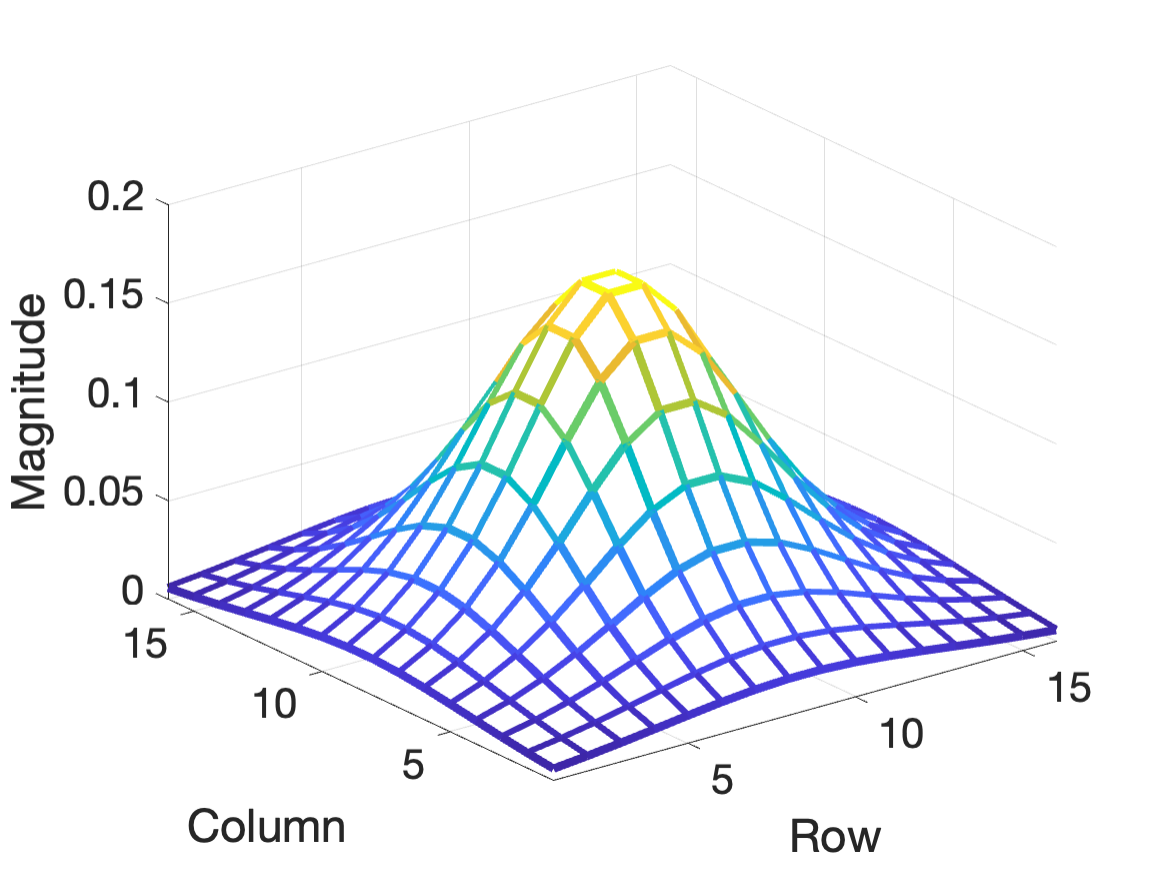}}
\caption{RIS PEM magnitude taper (normalized) as obtained from Friis-based modeling.}
\label{fig:RIS_Friis}
\end{figure}

We see in Fig. \ref{fig:RIS_FW} that the PEM excitation at the RIS from $\Tm_{\rm full.wave}$ has a smooth monotonic magnitude taper. This is qualitatively similar to that from the Friis-based modeling $\Tm_{\rm Friis}$ as seen in Fig. \ref{fig:RIS_Friis}. Quantitatively, the full-wave simulation yields the RIS power taper of $\unit[18.1]{dB}$ which is much smaller than the Friis-based RIS power taper of $\unit[29.5]{dB}$. Resultant far-field power radiation shows pencil beam peaks of $\unit[29.6]{dBi}$ and $\unit[29.2]{dBi}$, and peak side lobe levels of $\unit[5]{dB}$ and $\unit[-6.2]{dB}$, respectively. 


We performed the system level simulations for the parameters as in \cite[Section IV]{SPAWC}. Only the AMAF-RIS matrix $\Tm_{\rm Friis}$ and $\Tm_{\rm full.wave}$ were changed in two simulation runs. Also, the RF power increased by 3.52 dB to compensate for the increased path loss at 150 GHz as compared to 100 GHz in \cite[Section IV]{SPAWC}. We see in Fig. \ref{fig:Gemic_pointing_error} that, for the perfect beam pointing case, there is a close match between the information-theoretic communication rate cumulative distribution function (CDF) curves obtained from $\Tm_{\rm Friis}$ and $\Tm_{\rm full.wave}$. Due to the smaller PEM taper, the full-wave rate CDF curve is more sensitive to beam pointing errors than the Friis-based curve. Fig. \ref{fig:Gemic_digital_PS} shows the curves for discrete phase shifters at the RIS. Overall, it is clear that the Friis modeling can be relied upon to obtain first-order performance approximations, while for the ultimate performance evaluation, a more accurate full-wave simulation of the near-field AMAF-RIS transmission matrix is needed. The qualitative behavior of the two approaches is quite consistent, and the proposed AMAF-RIS scheme is likely to work even taking into account the non-negligible inter-element mutual couplings at the AMAF and the RIS.

\begin{figure}[h] 
\centerline{\includegraphics[width = 0.95\linewidth]{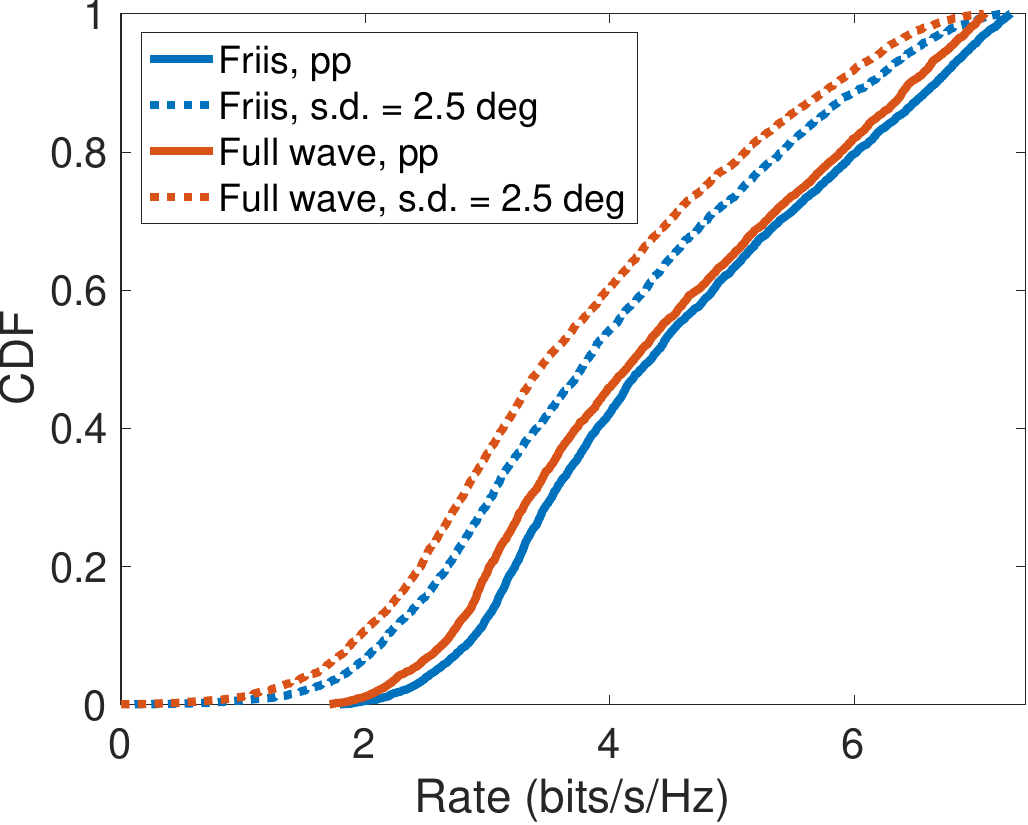}}
\caption{Rate CDF curves as obtained from $\Tm_{\rm Friis}$ and $\Tm_{\rm full.wave}$ with and without Gaussian beam pointing errors.}
\label{fig:Gemic_pointing_error}
\end{figure}

\begin{figure}[h] 
\centerline{\includegraphics[width = 0.45\textwidth]{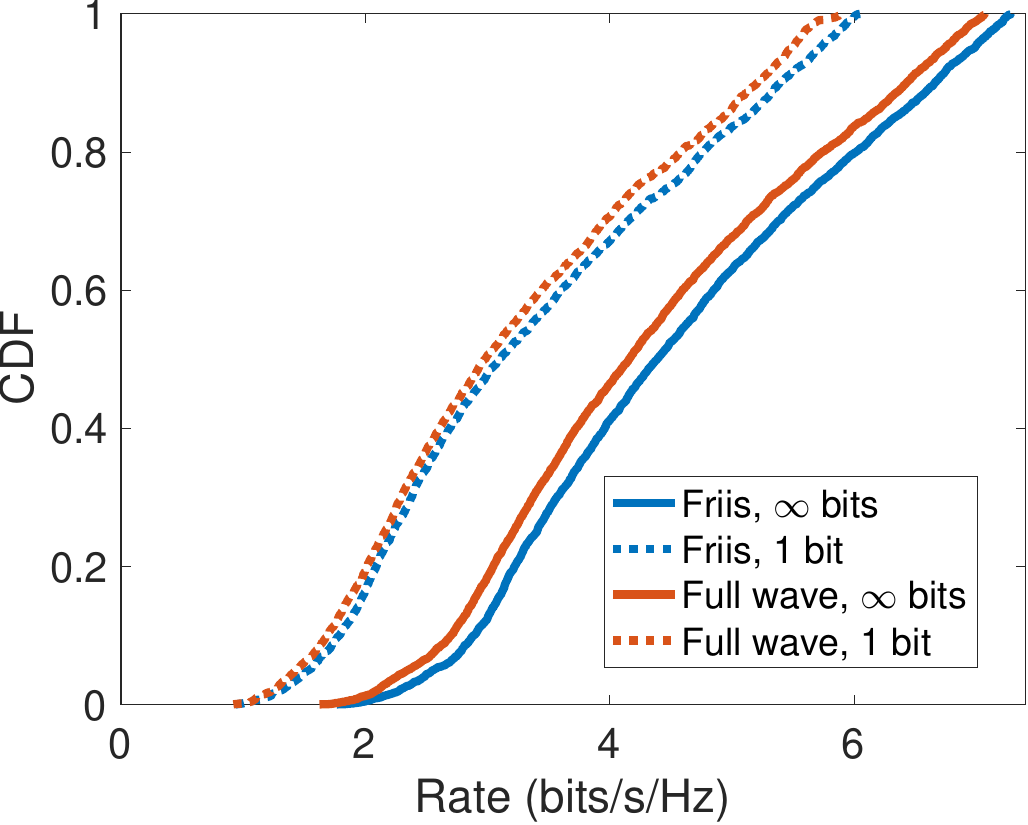}}
\caption{Rate CDF curves as obtained from $\Tm_{\rm Friis}$ and $\Tm_{\rm full.wave}$ with discrete phase shifters at the RIS.}
\label{fig:Gemic_digital_PS}
\end{figure}

\section{Conclusions}  
\label{sec:CONC}

We presented a comparative study of Friis-based modeling and full-wave simulations for the AMAF-RIS system. While Friis-based modeling provides a useful first-order approximation for system design, full-wave simulations capture more detailed effects such as mutual coupling between elements, etc., making them essential for a more accurate performance evaluation. Our results show a strong qualitative agreement between the two approaches, with full-wave simulations showing a smaller power taper and thus a higher sensitivity to beam pointing errors. Overall, the proposed AMAF-RIS architecture shows promise for real-world applications, with Friis-based models suitable for initial design stages and full-wave simulations required for more accurate performance evaluation. The study suggests ceiling the principal singular value at unity when it exceeds one in Friis-based models. Future research directions include full-wave simulations for stacked AMAF-RIS modules for multi-user MIMO, dual-polarized AMAF-RIS for serving twice the number of data-streams, and hardware-based validations in different frequency bands and different system settings. 

\section*{Acknowledgment}

The work was supported by BMBF Germany in the program of ``Souverän. Digital. Vernetzt.'' Joint Project 6G-RIC (Project IDs 16KISK030 and 16KISK024).

\bibliographystyle{IEEEtran}
\bibliography{P2-bibliography}
\end{document}